\documentclass[a4paper,11pt]{article}

\usepackage{jcappub} 

\usepackage[T1]{fontenc} 
\usepackage{newpxtext,newpxmath}
\usepackage{amsmath}
\usepackage{subcaption}
\usepackage{graphicx}

\usepackage{mathtools}
\DeclarePairedDelimiter\abs{\lvert}{\rvert}%
\DeclarePairedDelimiter\norm{\lVert}{\rVert}%

\makeatletter
\let\oldabs\abs
\def\abs{\@ifstar{\oldabs}{\oldabs*}}
\let\oldnorm\norm
\def\norm{\@ifstar{\oldnorm}{\oldnorm*}}
\makeatother
\def\L{{\mathcal L}}
\def\Tr{{\rm Tr}}

\def\Tr{{\rm Tr}}

\def\L{{\mathcal L}}
\def\vep{\varepsilon}

\newcommand{\Zstroke}{%
  \text{\ooalign{\hidewidth\raisebox{0.2ex}{--}\hidewidth\cr$Z$\cr}}%
}
\newcommand{\zstroke}{%
  \text{\ooalign{\hidewidth -\kern-.3em-\hidewidth\cr$z$\cr}}%
}

\newcommand\comment[1]{}
\newcommand\expect[1]{\left\langle #1 \right\rangle}

\makeatletter
\newcommand*\bigcdot{\mathpalette\bigcdot@{.5}}
\newcommand*\bigcdot@[2]{\mathbin{\vcenter{\hbox{\scalebox{#2}{$\m@th#1\bullet$}}}}}
\makeatother

\title{Warm Affine Inflation}

\author[a,b]{Mahmoud AlHallak}

\affiliation[a]{Al Ain University}
\affiliation[b]{Physics Dept., Damascus University, Damascus, Syria}

\emailAdd{mahmoud.alhallak@damascusuniversity.edu.sy}
\emailAdd{phy.halak@hotmail.com}

\abstract{The warm inflationary scenario is investigated in the context of affine gravity formalism. A general framework is provided for studying different single-field potentials.
\\Using the sphaleron mechanism we explain the continuous dissipation of the inflaton field into radiation, leading to the $\Gamma=\Gamma_0 T^3$ dissipation coefficient. The treatment is performed in the weak and strong dissipation limits. 
\\
We consider the quartic potential as a case study to provide a detailed study. Moreover, in this study, we discuss various constraints on inflationary models in general. 
\\
We compare the theoretical results of the quartic potential model within warm inflation with the observational constraints from Planck $2018$ and BICEP/Keck 2018, as presented by the tensor-to-scalar ratio, spectral index and the perturbation spectrum. }

\makeatletter
\gdef\@fpheader{}
\makeatother
\begin{document}
\maketitle
\flushbottom
\section{Introduction:}
Despite general relativity's achievements using its metric formalism, questions have always existed about its uniqueness and limitations as a gravitational theory. 
\\ The first attempts to examine the general relativity limitations appeared only four years after the theory was raised when Weyl began considering the possibility of modifying the theory to include higher-order invariants \cite{Weyl:1919fi}. It's interesting to know that Albert Einstein was one of those who began to examine assumptions within the theory. Einstein introduced the notion of affine connections, independent of the metric, and subsequently applied the variational principle to both. This approach later became known as the Palatini formalism of gravity \cite{Ferraris:1982,Palatini:1919}.
\\ 
It has been shown, however, that metric and palatini formalisms are equivalent in the case of minimal coupling between matter and gravity \cite{Wald:1984rg}. On the other hand, considering a modification of the gravitational theory, such as the non-minimal coupling to gravity or higher order of curvature terms shows curial differences between the two formalisms \cite{Ferraris:1992dx,Magnano:1993bd}.

Despite the fact that the connections in the Palatini formalism seem independent of the metric tensor, the theory as a whole cannot be considered completely affine, since the invariant volume can still be constructed using the square root of the metric tensor's determinant, and matter action depends only on the metric tensor, not on the affine connection. 

A pure affine formalism of gravity is proposed in \cite{Azri:2015diq}.
The model shows metric-free invariant action, where derivatives of scalar fields (kinetic part) enter dynamics along with curvature tensors, both appearing in the determinant of the invariant volume element. Rather than adding to the kinetic part, the non-derivative parts of the scalar field (potentials) are divided into it \cite{Azri:2015vta,Azri:2015diq,Azri:2018qcc}. 
\\Table \eqref{tab:formalisms_comparison} compares key quantities in the metric, Palatini, and affine formalisms.
\\ Similarly to Palatini formalism, affine and metric equivalencies appear under minimal coupling conditions.  Notably, significant findings emerge when non-minimal coupling is considered between matter and gravity. 

The affine formalism of gravity has been explored in various cosmological contexts \cite{Azri:2018xap,Azri:2020bzl,Azri:2018qcc,Azri:2020agc,Azri:2021dbr}, with particular attention given to the cold inflationary scenario \cite{Azri:2017uor,Azri:2018qux,Azri:2018gsz,Azri:2019ffj,Azri:2021uat}. 
\\

A well-established alternative to conventional (Cold) inflation is warm inflation \cite{Berera:1995ie,Berera:1995wh}, which involves thermally coupling an inflaton field to a radiation bath. Warm inflation exhibits mainly thermal fluctuations, with quantum fluctuations suppressed by high dissipation rates between the inflaton field and radiation sector. In contrast to cold inflation, the inflaton produces radiation continuously, eliminating the necessity for a reheating phase at the end of inflation. 
\\Both metric and palatini formulations have been used to study warm inflation with different inflationary potentials \cite{AlHallak:2021hwb, AlHallak:2022haa, Alhallak:2022szt, Samart:2021hgt, Kamali:2018ylz}.  Despite this, no studies have been conducted on warm inflation in the context of affine gravity. A major goal of this work is to investigate warm inflationary scenarios under the affine formalism of gravity and provide a basis for examining various models that are inspired by particle physics, such as \cite{Bezrukov:2007ep} or other theories, like \cite{AlHallak:2016wsa}.

The paper is organized as follows. In Sect. \ref{Sec:2}, we provide an overview of gravity's formalism based on affine geometry. Considering the minimal coupling between matter and gravity, we demonstrate the equivalence between the metric and affine formalisms. The field equations of the non-minimal coupling case are discussed later.
Sect. \ref{Sec:3} presents a general framework to study the warm inflationary scenario in the context of affine gravity. Equations of motion, slow roll approximation, and perturbation spectra are discussed. Finally, we summarize the equations of the main observables in both strong and weak dissipation limits.
In Sect. \ref{Sec:4}, a case study of quartic potential in the field is examined in the context of affine formalism. The main constraints  are thoroughly discussed along with the detailed treatment of the model. We end up with a summary and conclusion in Sect. \ref{sect:conclusion}.
\section{Review: Affine formalism of gravity}\label{Sec:2}
In this section, we examine the pure affine gravity formalism, reviewing its principles and concepts. We commence by exploring the minimal coupling between matter and gravity, demonstrating the equivalence between the metric and affine formalisms. \\Subsequently, we will introduce a non-minimal coupling case to illustrate the disparities between the two formalisms and their impact on inflationary outcomes.
\subsection{Minimal Coupling Case:}
As a starting point, we begin by considering the most straightforward manifestation of affine gravity:  
\begin{equation} \label{eq:min_SAG}
S_{AG}= \int d^4 x \sqrt{\norm {\tilde{K}_{\mu\nu}} } 
\end{equation}
where $ \sqrt{\norm {\tilde{K}_{\mu\nu}} }$ represents the invariant volume in the affine gravity with: 
\begin{equation} \label{eq:Kmunu}
K_{\mu\nu}=\Re_{\mu\nu}-\nabla_\mu \phi \nabla_\nu \phi   \quad \text{and} \quad \norm{\tilde{K}_{\mu\nu}}=\frac{\norm{K_{\mu\nu}}}{V^2(\phi)}
\end{equation}
 $\Re_{\mu\nu}$ is the Ricci tensor built out of the connection $\Gamma$ independently of the metric and, $V(\phi)$ is the potential of the scalar field $\phi$.

As we can see from the equations \eqref{eq:min_SAG} and,\eqref{eq:Kmunu} the curvature is tightly related to the space-time connection. In this context, it is important to note the similarity to the Palatini formalism, where the construction of the Riemann tensor also involves the connection. However, unlike the Palatini formalism, gravity is purely built upon the connection of $\Gamma$ in this scenario in which the invariant volume remains independent of the metric as well.
\\A notable feature of the above action is that it is singular at $V(\phi)=0$. Consequently, the model must always have nonzero potential energy. For $\phi_{min}=cons \neq 0$, $V(\phi_{min})\neq 0$ appears as a cosmological constant in the theory. 
\\

By taking the variation of the action \eqref{eq:min_SAG} with respect to the connection, we arrive at the field equations of this gravitational theory as,
\begin{equation} \label{eq:var_of_S}
\delta S = \int d^4 x \sqrt{\norm{\tilde{K_{\mu\nu}}}} (K^{-1})^{\mu\nu}\delta \Re_{\mu\nu}
\end{equation}
It can be shown that the variation of Ricci tensor $\Re_{\mu\nu}$ can be written in terms of of variation of connections $\Gamma^\alpha_{\mu\nu}$ as,  
\begin{equation}
\delta \Re_{\mu\nu}=\nabla_\alpha \delta \Gamma^\alpha_{\mu\nu}-\nabla_\nu \delta \Gamma^\alpha_{\mu\alpha}
\end{equation}
To isolate the variations, integration by parts is required, resulting in: 
\begin{equation}
\delta S = \int d^4 x \big[ \delta_\alpha^\nu \nabla_\beta (\sqrt {\tilde{\norm{K_{\mu\nu}}}} (K^{-1})^{\mu \beta})-\nabla_\alpha (\sqrt{\tilde{\norm{K_{\mu\nu}}}} (K^{-1})^{\mu \nu})\big]\delta \Gamma^\alpha_{\mu\nu}=0
\end{equation}
where $\delta_\alpha^\nu$ is the Kronecker delta, introduced here to help manipulate the indexes of $\delta \Gamma^\alpha_{\mu\nu}$ and factorize this term.
\\As $\delta \Gamma^\alpha_{\mu\nu}$ is symmetric in the indexes $\mu$ and $\nu$, this variation forces that the symmetric part of the expression inside the square brackets vanishes, i.e.,  
\begin{equation}\label{symmetric_variaition}
\frac{1}{2} \delta_\alpha^\nu  \nabla_\beta (\sqrt{\tilde{\norm{K_{\mu\nu}}}} (K^{-1})^{\mu \beta})+ \frac{1}{2} \delta_\alpha^\mu  \nabla_\beta (\sqrt{\tilde{\norm{K_{\mu\nu}}}} (K^{-1})^{\nu \beta}) - \nabla_\alpha (\sqrt{\tilde{\norm{K_{\mu\nu}}}} (K^{-1})^{\mu \nu})=0
\end{equation}
It can be shown that if the last equation \eqref{symmetric_variaition}  holds for arbitrary connections, then,
\begin{equation}\label{min_nabla_K}
\nabla_\alpha (\sqrt{\tilde{\norm{K_{\mu\nu}}}} (K^{-1})^{\mu \nu})=0
\end{equation}
as well as, 
\begin{equation}
\nabla_\alpha  ((K^{-1})^{\mu \nu})=0
\end{equation}
We can conclude from the last equation that the affine connection that has been taken arbitrary but symmetric obeys a constraint that reduces it to a levi-civita connection of an invertible metric  $\tilde{g}_{\mu\nu}$. In other words, this tensor appears as a solution to the equation by setting : 
\begin{equation}
\frac{\sqrt{\norm{K_{\mu\nu}}}}{V(\phi)} (K^{-1})^{\alpha \beta}=M_{pl}^2 \sqrt{\norm{\tilde{g}_{\mu\nu}}}(\tilde{g}^{-1})^{\alpha\beta}
\end{equation}
which leads to:
\begin{equation}\label{eq:Kwithgtilde}
K_{\mu\nu}=M_{pl}^2 V(\phi) \tilde{g}_{\mu\nu}
\end{equation}
\begin{equation}
\nabla_\alpha  (\tilde{g}_{\mu \nu})=0
\end{equation}
Using equations \eqref{eq:Kmunu} and \eqref{eq:Kwithgtilde} we can find Ricci tensor as: 
\begin{equation}
\Re_{\mu\nu}= \nabla_\mu \phi \nabla_\nu + M^2_{pl} V(\phi) \tilde{g_{\mu\nu}}
\end{equation}
and Ricci scalar as:
\begin{equation}
\Re= \tilde{g}^{\mu\nu}\Re_{\mu\nu}= \nabla^\alpha \phi \nabla_\alpha \phi + 4 M^2_{pl} V(\phi) 
\end{equation}
Now we can write Einstein equations which appears as: 
\begin{equation}
\Re_{\mu \nu}-\frac{1}{2}\tilde{g}_{\mu\nu}\Re = \nabla_\mu \phi \nabla_\nu \phi - \frac{1}{2} \tilde{g}_{\mu \nu} \nabla^\alpha \phi \nabla_\alpha \phi-M_{pl}^2 V(\phi)\tilde{g}_{\mu\nu}
\end{equation}
Considering the last equation, we can conclude that the affine formalism of gravity is equivalent to the metric formalism of standard general relativity in the minimal coupling case.

\begin{table}
\renewcommand{\arraystretch}{2}
    \centering
    \begin{tabular}{|c|c|c|c|} \hline 
         &  Metric&  Palatini& Affine\\ \hline 
         Ricci Tensor&  $R_{\mu\nu}(\partial^2 g)$&  $R_{\mu\nu}(\Gamma)$& $R_{\mu\nu}(\Gamma)$\\ \hline 
         Invariant Volume Element&  $d^4x \sqrt{\norm{g}}$&  $d^4x \sqrt{\norm{g}}$& $d^4x \sqrt{\norm{K}}$\\ \hline 
         Matter Action&  $S_M(\psi,g_{\mu\nu})$&  $S_M(\psi,g_{\mu\nu})$& included in $\norm{K(\Gamma,\psi)}$\\ \hline
    \end{tabular}
    \caption{Comparison between key quantities in the metric, Palatini, and affine formalisms. Here, $\psi$ represents the matter fields, $||\text{g}||$ denotes the determinant of the metric tensor $g_{\mu\nu}$, and $||\text{K}||$ represents the determinant of the tensor $K_{\mu\nu}$. }
    \label{tab:formalisms_comparison}
\end{table}
\subsection{Non-minimal Coupling Case:}
In this section we shall treat the case of non-minimal coupling between the matter represented by a scalar field $\phi$ and the curvature represented by Ricci tensor $R_{\mu \nu}$. 
\\Concerning this situation the action will take the form: 
\begin{equation}\label{eq:SAG_NMC}
S_{AG}= \int d^4 x  \sqrt{\frac{\norm{f(\phi) \Re_{\mu\nu}(\Gamma)-\partial_\mu \phi \partial_\nu \phi}}{V^2(\phi)}}
\end{equation}
As a result, equation \eqref{min_nabla_K} will take the form:
\begin{equation}
\nabla_\alpha \bigg\{F(\phi) \sqrt{\norm{f(\phi)\Re_{\mu\nu}-\partial_\mu \phi \partial_\nu \phi}}\big(f(\phi) \Re^{\mu\nu}-\partial^\mu \phi \partial^\nu \phi \big)\bigg\}=0
\end{equation}
where $F(\phi) \equiv \frac{f(\phi)}{V(\phi)}$.
\\
Einstein's equations are given now as: 
\begin{equation}
 f(\phi) G_{\mu\nu}=\partial_\mu \phi \partial_\nu \phi -\frac{1}{2}g_{\mu\nu}\partial_\alpha \phi \partial^\alpha \phi -g_{\mu\nu}F(\phi)
\end{equation}
Rearranging the last equation results in: 
\begin{equation}
 G_{\mu\nu}=T_{\mu\nu}^\phi + T_{\mu\nu}^{AG}
\end{equation}
where: 
\begin{equation}
T_{\mu\nu}^\phi=\partial_\mu \phi \partial_\nu \phi -\frac{1}{2}g_{\mu\nu}\partial_\alpha \phi \partial^\alpha \phi -g_{\mu\nu}V(\phi)
\end{equation}
\begin{equation}
T_{\mu\nu}^{AG}=(f(\phi)-1)\big[g_{\mu\nu}F^{-1}(\phi)-G_{\mu\nu}\big]
\end{equation}
Obviously, the last quantity vanishes for $f(\phi)=1$ – the minimal coupling case. 

In order to establish a general framework for the model, we keep the function $f(\phi)$ undefined, as its specific definition can vary depending on the particular model under investigation. One example of a non-minimal coupling term is the quadratic term involving the scalar field. When examining a cosine potential with an inflaton field that respects the shift symmetry of the potential, the function $f(\phi)$ can take on a comparable periodic form \cite{Ferreira:2018nav,AlHallak:2022gbv,AlHallak:2023wsx}. In \cite{AlHallak:2016wsa,AlHallak:2021sic}, the authors show that the NMC term can be expressed exponentially in terms of the scalar field.
\section{Warming up Affine inflation:}\label{Sec:3}
In this section, our focus will be on exploring the warm inflationary scenario within the framework of affine gravity.  
\subsection{Equations of Motion:}
The main characteristic of warm inflation is that, unlike cold inflation, thermal radiation is continuously produced throughout the inflationary period.  
\\Let's start with the general action as: 
\begin{equation}
S=S_{AG}+S_{rad}+S_{Int}
\end{equation}
Where $S_{AG}$ is the action \eqref{eq:SAG_NMC}, $S_{rad}$ is the action of the radiation produced in the inflationary epoch and $S_{Int}$ represents the interaction between the scalar field $\phi$ and radiation.

We assume the transferring of energy between the scalar field and radiation through the term:
\begin{equation}\label{Jmu}
J_\mu=\Upsilon u^\nu \nabla_\nu \phi \nabla_\mu \phi
\end{equation}
Where $u^\nu$  is the four-velocity vector $(1,0,0,0)$, and $\Upsilon$ is the dissipative coefficient that is a function of both the scalar field and temperature in general.
\\ As for radiation and inflaton fields, the continuity equations are given as follows: 
\begin{equation}
\nabla^\mu T_{\mu \nu}^{rad}=J_\nu
\end{equation}
\begin{equation}
\nabla^\mu (T_{\mu \nu}^{\phi}+T_{\mu \nu}^{AG})=-J_\nu
\end{equation}

A notable feature of affine gravity is that we can recover a minimal coupling scenario without the need for conformal transformation. This can be accomplished by simply redefining the scalar field. 
\\Looking at action \eqref{eq:SAG_NMC}, we can recover the minimal coupling case by redefining the scalar field as:
\begin{equation}\label{eq:dchi}
\frac{d\chi}{d\phi}=\frac{1}{\sqrt{f(\phi)}}
\end{equation}
The action \eqref{eq:SAG_NMC} now takes the form, 
\begin{equation}\label{SAG_NMC}
S_{AG}= \int d^4 x  \sqrt{\frac{\norm{ \Re_{\mu\nu}(\Gamma)-\partial_\mu \chi \partial_\nu \chi}}{U^2(\chi)}}
\end{equation}
Where:
\begin{equation}
U(\chi(\phi)) \equiv \frac{V(\phi)}{F^2(\phi)}
\end{equation}
It is noteworthy that the affine formalism is defined only by one unique frame, unlike the metric and Palatini formalisms \cite{Azri:2017uor,Azri:2018gsz}.

In accordance with assumption \eqref{Jmu}, the continuity equations will take the following forms:  
\begin{equation}\label{nablaT_chi}
\nabla^\mu T_{\mu \nu}^{\chi}=- \Zstroke J_\nu(\chi)
\end{equation}
\begin{equation}\label{nablaT_rad_chi}
\nabla^\mu T_{\mu \nu}^{rad}=- \Zstroke J_\nu(\chi)
\end{equation}
where $\Zstroke=f(\phi)$. 
\\Under the palatini formalism, dissipation terms are directly expressed in terms of $\chi$ as $J_\mu=\Upsilon u^\nu \nabla_\nu \chi \nabla_\mu \chi$ in \cite{Kamali:2018ylz}.
Therefore, in equations \eqref{nablaT_chi} and \eqref{nablaT_rad_chi}, the term $Z$ is assumed to be unity on the right side. However, this work proceeds in accordance with assumption \eqref{Jmu} by considering the dissipation of the physical field $\phi$ rather than rescaled $\chi$. 

Friemann equations are given by the equations:
\begin{equation}\label{friendmn_1}
H^2=\frac{1}{3M_{pl}^2}(\rho_\chi + \rho_{rad})
\end{equation}
\begin{equation}\label{friendmn_2}
\dot{H}=-\frac{1}{2M_{pl}^2}(\rho_\chi + \rho_{rad}+P_\chi + P_{rad})
\end{equation}
where:
\begin{equation}\label{}
\rho_\chi=\frac{1}{2}\dot{\chi}^2+U(\chi)
\end{equation}
\begin{equation}\label{}
P_\chi=\frac{1}{2}\dot{\chi}^2-U(\chi)
\end{equation}
And $H$ is Hubble parameter. 
\\ The equations of motion for the radiation and inflaton fields are now expressed as follows:
\begin{equation}\label{EOM_rad}
\dot{\rho}_{rad}+4H\rho_{rad}=\Upsilon \Zstroke \dot{\chi}^2
\end{equation}
\begin{equation}\label{EOF_chi_chi}
\ddot{\chi}+3H\dot{\chi}+U_\chi=-\Upsilon \Zstroke \dot{\chi}
\end{equation}

\subsection{Slow Roll Approximation:}
A slow-roll approximation,
\begin{equation}
\label{slow_appr_1}
\ddot{\chi} \ll H \dot{\chi}  ,  \dot{\rho_\gamma} \ll H \rho_\gamma
\end{equation}
is usually used to calculate inflationary solutions to equations (\ref{friendmn_1}),(\ref{friendmn_2}) , (\ref{EOM_rad}), and (\ref{EOF_chi_chi}). 
When applying the slow-roll approximation to the exact equations, the highest order terms typically are omitted, 
During the slow-roll regime, eqs. (\ref{friendmn_1}),(\ref{friendmn_2}) , (\ref{EOM_rad}),  and (\ref{EOF_chi_chi}) read:
\begin{equation}
\label{H_slow_roll}
H^2 \simeq \frac{1}{3} U 
\end{equation}\begin{equation}
\label{dotH_slow_roll}
\dot{H} = -\frac{1}{2} (1+\Zstroke Q) \dot{\chi}^2
\end{equation}
\begin{equation}
\label{rho_gamma_slow_roll}
\rho_\gamma \simeq \frac{3}{4} \Zstroke Q \dot{\chi}^2
\end{equation}
\begin{equation}
\label{chi_gamma_slow_roll}
\dot{\chi} = - \frac{U_\chi}{3 H (1+ \Zstroke Q)}
\end{equation}where, 
\begin{equation}
\label{Q}
Q= \frac{\Upsilon}{ 3 H}
\end{equation}
The parameter $Q$ represents the relative strength of the thermal damping compared to the expansion damping.  Compared to cold inflation, warm inflation has relatively large values of Q, while cold inflation neglects Q entirely.

In order to define model-dependent slow roll parameters, we follow the standard procedure starting from the model-independent parameters. 
\par (a) For a quasi-exponential expansion of the universe, the first model-independent parameter is $-\frac{\dot{H}}{H^2}\ll 1$. Using equations \eqref{H_slow_roll} and (\ref{dotH_slow_roll}) we can find: 
\begin{equation}
\label{H_epsilon}
-\frac{\dot{H}}{H^2} = \frac{\varepsilon}{(1+\Zstroke Q)} \ll 1
\end{equation}
where 
\begin{equation}
\label{epsilon}
\varepsilon= \frac{1}{2} \bigg(\frac{U_\chi}{U}\bigg)^2
\end{equation}
is the first model-dependent slow roll parameter.
\\The condition (\ref{H_epsilon})  imposes,
\begin{equation}
\varepsilon \ll (1+ \Zstroke Q) 
\end{equation}
This constrain  is slightly different from the standard warm inflation, where the condition reads ${\varepsilon}\ll{(1+ Q)} $.
\par (b)The second important parameter is $|{\frac{\ddot{\chi}}{H\dot{\chi}}}|\ll 1$, which is crucial to obtaining the necessary number of e-folds. Using Eqs (\ref{H_slow_roll} ) , (\ref{dotH_slow_roll}) and (\ref{chi_gamma_slow_roll}), we find,
\begin{equation}
\label{phidotdot}
\frac{\ddot{\chi}}{H\dot{\chi}} = - \frac{U_{\chi\chi}}{U (1+\Zstroke Q)}-\frac{\dot{H}}{H^2 (1+ \Zstroke Q)} +\frac{\Zstroke  Q U_\chi \Upsilon_\chi }{U \Upsilon  (1+ \Zstroke Q)^2}+\frac{\Zstroke  Q U_\chi \Zstroke_\chi }{U \Zstroke (1+ \Zstroke Q)^2}
\end{equation}
The first term of  eq (\ref{phidotdot}) defines the second slow-roll parameter as,
\begin{equation}
\label{eta}
\eta \equiv \frac{U_{\chi \chi}}{U}
\end{equation}
with the constrain, 
\begin{equation}
\frac{\abs{\eta}}{(1+ \Zstroke Q)} \ll 1 
\end{equation}
The second term of eq (\ref{phidotdot}) is related to the first slow-roll parameter $\varepsilon$ \eqref{H_epsilon}, while the third and fourth relate to the third and fourth slow-roll parameters, respectively; we will explore these next.
\par (c)  In view of the quasi-stable radiation production (\ref{slow_appr_1}), we can define slow-roll parameters by taking the differentiation of equation (\ref{rho_gamma_slow_roll}),
\begin{equation}
\label{dot_rho_gamma}
\frac{\dot{\rho_\gamma}} {  H \rho_\gamma} = -\frac{\Upsilon_\chi U_\chi}{\Upsilon U (1+ \Zstroke Q)} - \frac{\Zstroke_\chi U_\chi}{\Zstroke U (1+ \Zstroke Q)}-\frac{\dot{H}}{H^2} + {\frac{2 \ddot{\chi}}{H\dot{\chi}}} \ll 1
\end{equation}
The sufficient condition for establishing Eq.(\ref{dot_rho_gamma}) is the first and second term $\ll 1$. then, the third and fourth slow-roll parameters are expressed as,
\begin{equation}
\label{sigma_parameter}
\sigma \equiv \frac{\Zstroke_\chi U_\chi}{\Zstroke U}
\end{equation}
and 
\begin{equation}
\label{beta}
\beta \equiv \frac{\Upsilon_\chi U_\chi}{\Upsilon U}
\end{equation}
As a result, we can write (\ref{phidotdot})  as, 
\begin{equation}
\label{chi_dotdot}
{\frac{\ddot{\chi}}{H\dot{\chi}}} = - \frac{\eta}{(1+\Zstroke Q )}+\frac{\varepsilon}{(1+ \Zstroke Q)^2}+\frac{\Zstroke Q \beta}{(1+\Zstroke Q)^2}+\frac{\Zstroke Q \sigma}{(1+\Zstroke Q)^2}
\end{equation}
The third term of eq \eqref{chi_dotdot} can be expressed as $\frac{\beta}{(1+\Zstroke Q)}-\frac{\beta}{(1+\Zstroke Q)^2}$. Even though condition ${\frac{\ddot{\chi}}{H\dot{\chi}}}\ll 1$ requires the subtraction to be less than one, we assume that,
\begin{equation}
    \frac{\abs{\beta}}{(1+\Zstroke Q)}\ll 1
\end{equation}
to meet condition \eqref{dot_rho_gamma}.
 \\Following the same argument regarding the fourth term of eq \eqref{chi_dotdot}, we find the constraint,
 \begin{equation}
    \frac{\abs{\sigma}}{(1+\Zstroke Q)}\ll 1
\end{equation}
The same quantities can be formally defined for the field $\phi$ but are not of physical meaning. However, they are easier to compute and are related to the physical slow roll parameters via
\begin{equation}
\varepsilon=\Zstroke \varepsilon_\phi
\end{equation}
\begin{equation}
\eta=\Zstroke \eta_\phi + sgn (U^\prime) {\Zstroke^\prime} \sqrt{\frac{\varepsilon_\phi}{2}}
\end{equation}
\begin{equation}
\beta=\Zstroke \beta_\phi 
\end{equation}
\begin{equation}
\sigma=\Zstroke \sigma_\phi
\end{equation}
where prime means $d/d\phi$
\subsection{Perturbation Spectra:}
It is a common view that the seeds of the large-scale structure and the observed anisotropy of the cosmological microwave background are generated within the inflationary epoch through the vacuum fluctuations of inflaton field \cite{Clarkson:2003af, Liddle:2000cg, Weinberg:2008zzc}. The generation of seeds in warm inflation, however, is dominated by larger thermal fluctuations rather than quantum fluctuations\cite{Berera:1995ie, Berera:1995wh,Berera:1999ws}. 
\\
The primordial power spectrum of warm inflation at the horizon crossing is altered as \cite{Bastero-Gil:2016qru} ,
\begin{equation}
\Delta_R^2= \frac{U_\ast (1+\Zstroke Q_\ast)^2}{24 \pi^2 \Zstroke \varepsilon_{\phi\ast}} \big(1+ 2 n_\ast + \omega_\ast \big)
\end{equation}
where $n=\big(\exp H/T -1 \big)^{-1}$
is the Bose-Einstein statistical function, $\omega=\frac{T}{H}\frac{2\sqrt{3}\pi \Zstroke Q}{\sqrt{3+4\pi \Zstroke Q}}$ , and $\ast$ denotes the parameters at the horizon crossing.
\\
The general formula for the temperature $T$ can be derived by combining Equations (\ref{H_slow_roll}), (\ref{rho_gamma_slow_roll}), and (\ref{chi_gamma_slow_roll}) in the context of the slow-roll approximation, along with the following equations:
\begin{equation}
\label{eq:rho_R}
\rho_R=\alpha T^4 ,  P_R=\rho_R/3
\end{equation}
as,
\begin{equation}\label{eq:genral_T}
T=\bigg(\frac{U}{2\alpha}\frac{\Zstroke Q_\ast}{(1+\Zstroke Q_\ast)^2}\varepsilon_\ast \bigg)^{1/4}
\end{equation}
The scalar spectral index, commonly denoted as $n_s$, is:
\begin{equation}\label{eq:genral_ns}
n_s=\lim_{k -> k_\ast}\frac{d \ln \Delta_R^2(k/k_\ast)}{d \ln(k/k_\ast)}
\end{equation} 
The tensor-to-scalar perturbation ratio, $r$ is defined as, 
\begin{equation}\label{eq:genral_r}
r=\frac{\Delta_T^2}{\Delta_R^2}
\end{equation} 
Where $\Delta_T^2$ is the power spectrum of the tensor perturbation, \begin{equation}
\Delta_T^2=2 H^2/\pi^2
\end{equation}
As a next step, we discuss the observables under weak and strong dissipation. 
\\

A complete analysis of the model requires accurate identification of the dissipation coefficient function $\Upsilon$.
As a general rule,  $\Upsilon$ depends on both the inflaton field $\phi$ and the temperature T. However, in the analyses of the reheating stage within the framework of cold inflation, the dissipation coefficient $\Upsilon$ is often treated as an arbitrary parameter \cite{Albrecht:1982mp}.
Precise formulae for the dissipation coefficient $\Upsilon$ were acquired from analyzing particle production \cite{Abbott:1982hn,Hall:2004zr,Hall:2003zp}.
\\In this study, we consider a  specific mechanism of producing radiation known as a sphaleron heating mechanism \cite{Manton:1983nd,Arnold:1987mh,Arnold:1987zg}. The sphaleron mechanism, originally a well-known mechanism in particle physics , is now being explored in the context of warm inflation \cite{Mirbabayi:2022cbt}.
\\ Following \cite{Mirbabayi:2022cbt}, the inflaton field $\phi$  is considered as an axion field and the source of particle production is its coupling to the Yang-Mills theory:
\begin{equation}\label{DL}
\Delta \L = \frac{\alpha\phi}{16 \pi f} \Tr[G_{\mu\nu} \tilde G^{\mu\nu}],
\end{equation}
where $\alpha= \frac{g_{\rm YM}^2}{4\pi}$ and $\tilde G^{\mu\nu}= \vep^{\mu\nu\alpha\beta}G_{\alpha\beta}$. 
\\After considering sphelaron mechanism:
\begin{equation}\label{eq:sphaleron_1}  
\frac{\alpha}{16\pi} \expect{\Tr[G_{\mu\nu} \tilde G^{\mu\nu}]} \approx \frac{\Gamma}{2T} \frac{\dot\phi}{f},
\end{equation}
where $\Gamma$ is called the sphaleron rate.
\\
Equation \eqref{eq:sphaleron_1} in turn implies a friction coefficient $\gamma = \Gamma/(2Tf^2)$ in the $\phi$ equation of motion.
\\ Based on \cite{Arnold:1996dy} , the authors argued that: 
\begin{equation}\label{eq2}
\Gamma\propto \alpha^5 T^4
\end{equation}
As a result of equations \eqref{eq:sphaleron_1} and \eqref{eq2},one can find that $\Upsilon(T)\approx \frac{\alpha^5}{f^2} T^3$.
\\ In the following sections, we consider the general function of $\Upsilon$ to be denoted as:
\begin{equation} \label{eq:UpsilonEq}
\Upsilon(T)=\Upsilon_0 T^3
\end{equation}
\subsubsection{Weak Dissipation Limit}
A weak dissipation limit is realized by $Z.Q\ll 1$. The formulas of temperature, spectral index, and tensor to scalar ratio can therefore be found using equations \eqref{eq:genral_T},\eqref{eq:genral_ns}, \eqref{eq:genral_r}, and \eqref{eq:UpsilonEq}as,  
\begin{equation}
\label{eq:Tweak}
T = \bigg(\frac{U_\phi^2 \Zstroke^2 \Upsilon_0}{36 \alpha H^3} \bigg)
\end{equation}
\begin{equation}
\label{eq:ns_weak}
n_s-1 = \Zstroke (-6 \varepsilon_\phi + 2\eta_\phi+sgn(U^\prime)\sqrt{2}\frac{\Zstroke^\prime}{\Zstroke}\sqrt{\varepsilon_\phi})+\frac{\pi \Zstroke^4 \Upsilon_0 Q}{4 \alpha}\varepsilon_\phi(15 \varepsilon_\phi - 2 \eta_\phi- sgn(U^\prime)\sqrt{2}\frac{\Zstroke^\prime}{\Zstroke}\sqrt{\varepsilon_\phi}-9 \beta_\phi- 9 \sigma_\phi) 
\end{equation}
\begin{equation}
\label{eq:r_weak}
r=\frac{\Zstroke H}{2 T}16 \varepsilon_\phi
\end{equation}
Equation \eqref{eq:ns_weak} indicates that the difference between the spectral index $n_s$ in a cold inflationary scenario and a warm inflationary scenario arises from the fourth term. As long as the condition $Q\ll 1$ holds, this term is only significant when $Z^2.\Upsilon_0 \gg 1$, otherwise, it becomes negligible, and the numerical values of the spectral index will be as it is in a cold inflationary scenario. 
\\
With respect to the tensor-to-scalar ratio, the quantity $\frac{H}{2 T}$ is $>1$ since the temperature in warm inflationary scenarios is always $T > H$. Therefore, one should expect smaller values of $r$ in warm inflationary scenarios than the cold ones.
\subsubsection{Strong Dissipation Limit}
In the strong dissipation limit where $Z.Q\gg 1$ ,  
we can find the following relations: 
\begin{equation}
\label{eq:Tstrong}
T = \bigg(\frac{U_\phi^2 }{4 \alpha \Upsilon_0 H} \bigg)^{1/7}
\end{equation}
\begin{equation}
n_s-1=\frac{1}{ Q}(-\frac{9}{4}\varepsilon_\phi+\frac{3}{2}\eta_\phi+\frac{3}{2}sgn(U^\prime)\frac{\Zstroke^\prime}{\Zstroke}\sqrt{\frac{\varepsilon_\phi}{2}}) -\frac{9}{4}\beta_\phi -\frac{9}{4}\sigma_\phi) 
\end{equation}
\begin{equation}
r=\frac{H}{T}(\frac{16\varepsilon_\phi}{\sqrt{3 \pi} (\Zstroke^{3/2} Q)^{5/2}})
\end{equation}
\section{Case Study: Quartic Potential:}\label{Sec:4}
In this section, we study a specific model represented by quartic potential. A study is conducted both with strong and weak dissipation limits. 
\\ Our goal is to provide a general framework that can be used to determine the important observables when warm inflation is present, while at the same time respecting the needed constraints for achieving successful inflation.

We consider the potential: 
\begin{equation}\label{potential_phi}
V(\phi)= V_0 + \lambda \phi^4
\end{equation}
As we discussed earlier, solving equation \eqref{eq:min_SAG} at $\phi=0$ can result in a singularity. We avoid this singularity by adding the cosmological constant $V_0$ to the quartic term in the equation \eqref{potential_phi}. 

 Further, we consider the non-minimal coupling NMC function  \eqref{NMC_term} that is produced at one-loop order in the interacting theory as, 
\begin{equation}\label{NMC_term}
f(\phi)=1+ \xi \phi^2
\end{equation}

By considering the NMC term \eqref{NMC_term}, it is possible to find analytical relationships between $\phi$ and $\chi$ by solving equation \eqref{eq:dchi}, 
\begin{equation}
\phi = \frac{\sinh \left(\chi \sqrt{\xi}\right)}{\sqrt{\xi}}
\end{equation}
This leads us to use either the formulas of the observables under study in terms of $\phi$ or in terms of $\chi$. However, the $\phi$ equations provide a general framework since it may not be possible to find the analytical relationship between the two fields in all cases as it is dependent on the type of NMC term. 

\begin{figure}[t!]
  \begin{subfigure}{\linewidth}
  \includegraphics[scale=0.5]{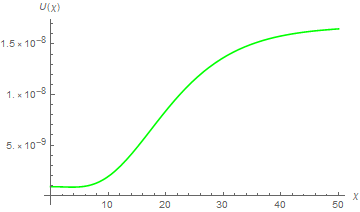}{(a)}\hfill
  \includegraphics[width=.45\linewidth]{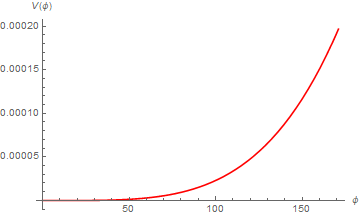}{(b)}\hfill
  \end{subfigure}\par\medskip
  \caption{(left panel) Ratio of the potential $U$ as a function of  field $\chi$. (right panel) the potential $V$ as a fuction $\phi$. The graphs plotted with $\xi = 3.7\times 10^{-3}$, $\Upsilon_0 = 4\times 10^4$, $\lambda = 2.3\times 10^{-13}$, $\alpha = 75$, and $V_0 = 9.0\times 10^{-9}$.}\label{fig:V_vs_U}
\end{figure}

\begin{figure}\centering 
  \includegraphics[scale=0.7]{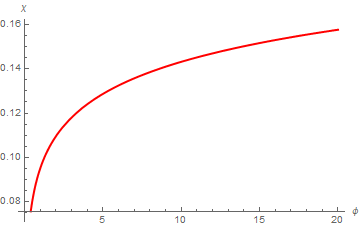}
  \caption{Plot of field $\chi$ as a function of $\phi$ with $\xi =2.3\times 10^3$.}\label{fig:phivschi}
\end{figure}

According to field $\chi$, the potential \eqref{potential_phi} can be expressed as,
\begin{equation}
U(\chi)=V_0 + \frac{\lambda \sinh ^4\left(\chi \sqrt{\xi}\right)}{\xi^2}
\end{equation}
Figure \eqref{fig:V_vs_U} represents the potentials $V$ and $U$ as a kind of comparison. According to the figure, the inflation type we are seeing is large field inflation, in which the inflaton field at the end of inflation is less than the field at the crossing horizon, and the field rolls downward from right to left. Nevertheless, it is important to consider whether the field $\chi$ has an increasing or decreasing relationship with $\phi$. In Figure \eqref{fig:phivschi}, the field $\phi$ is shown as an increasing function, which allows us to proceed safely with the large field inflation scenario in terms of both fields $\phi$ and $\chi$. 

Next, we examine the observables $n_s$, $r$, and $A_s$ as well as the number of e-foldings, which is an essential quantity in solving the flatness problem.
\\ A summary of the main constraints we should keep in mind when investigating observables can be found in Table \eqref{tab:constraints}. It was challenging to find the observables that complied with all theoretical constraints along with the wide field of parameter spaces. 
\begin{table}[]
\renewcommand{\arraystretch}{2}
\begin{tabular}{|llll|}
\hline
\multicolumn{4}{|c|}{Horizon Crossing}                                                                                                                                                                                  \\ \hline
\multicolumn{1}{|l|}{$n_{s\ast} = 0.9649 \pm 0.0042$}             & \multicolumn{1}{l|}{$r_{\ast} < 0.056$}                 & \multicolumn{1}{l|}{$Z(\phi_\ast)\bigcdot Q(\phi_\ast) \ll / \gg 1$}             & $\omega(\phi_\ast)\ll / \gg 1$                     \\ \hline
\multicolumn{1}{|l|}{$T_\ast > H_\ast$}                           & \multicolumn{1}{l|}{${A_s}_\ast = 2.196_{-0.060}^{+0.051}$}     & \multicolumn{2}{c|}{$\varepsilon_{\ast},\eta_{\ast},\beta_{\ast} and \sigma_{\ast} < (1+Z \bigcdot Q)_{\ast}$}                                              \\ \hline
\multicolumn{4}{|c|}{By End of Inflation}                                                                                                                                                                                   \\ \hline
\multicolumn{1}{|l|}{$Z(\phi_{end}) \bigcdot Q(\phi_{end})\ll / \gg 1$} & \multicolumn{2}{l|}{$\omega(\phi_{end})\ll / \gg 1$} & \multicolumn{1}{l|}{$T_{end}> H_{end}$}                             \\ \hline
\multicolumn{1}{|l|}{$N \in [40-60]$}                   & \multicolumn{3}{c|}{$\varepsilon_{end},\eta_{end},\beta_{end} or\sigma_{end} = (1+Z\bigcdot Q)_{end}$}                                                                          \\ \hline
\end{tabular}
\caption{The main constraints to be considered as one proceeds in the warm inflation scenario. In a strong limit, $Z.Q$ and $\omega$ are $\gg1$; in a weak limit, they are smaller than 1. regarding the  value of ns and r is obtained according to ref \cite{Planck:2018jri} at $68 \% CL$ after combinig the reustls of (Planck TT,TE,EE+lowE+lensing), bound of $r$ is for $95 \% CL$ with the combination of (Planck TT,TE,EE +lowE+lensing+BK15)\cite{Planck:2018jri}
.however, border ranges can be found in figure \eqref{fig:weak_countors} for different combined observations with different confidence levels.$A_s$ is obtained at $68 \% CL$ for Planck+WP \cite{Planck:2013pxb}.}
\label{tab:constraints}
\end{table}
\subsection{ Comparative analysis with observation:}
\subsubsection{Strong Dissipation Limit:}
In this limit, parameter spaces are extensively scanned. Because of the large number of parameters in the model, after finding an accessible observation point, further analysis is done by scanning specific parameters and fixing the others to determine the effect of each parameter on the model's output.

\begin{figure}
    \centering
    \includegraphics[width=1.0\linewidth]{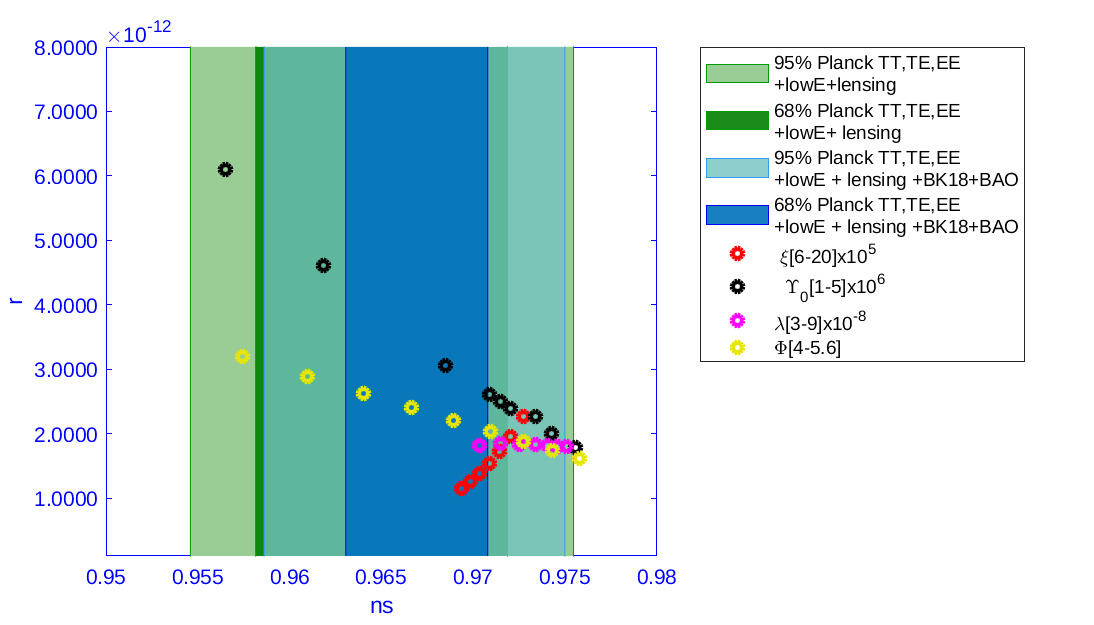}
        \caption{A plot of $r$ and $n_s$ under the strong dissipation limit in the case of $\Gamma=\Gamma_0 T^3$. The red dots correspond to the parameters given in equation \eqref{Eq:xi_scane}. The parameters of black dots are described by equation \eqref{eq:upsilon_scan}. A yellow point's parameters are given by equation \eqref{eq:phi_scan}. }
    \label{fig: strong_countors}
\end{figure}

Starting with NMC parameter $\xi$, we scan it in the range $[6-20]\times10^5$ with other parameters fixed as follows:
\begin{equation}\label{Eq:xi_scane}
\Upsilon_0=4 \times 10^6, \lambda=1\times10^{-7}, V_0=1\times10^{-12},\text{and} \phi=5.9. 
\end{equation}
As shown in figure \eqref{fig: strong_countors}, increasing parameter $\xi$ leads to a decrease in both the spectral index $n_s$ and the tensor to a scalar ratio $r$,  which exhibited in ranges of $[0.973-0.966]$ and $[22.7-6.72]\times 10^{-13}$, respectively. We can see from figure \eqref{fig: strong_countors} that these results are in the $68\%$ and $95\%$  confidence levels for Planck TT,TE,EE+LowE+lensing+BK18+BAO combinations. 
\\Additionally, the curvature perturbation spectrum lies between $[8.28-2.51]\times10^{-9}$ , which includes the observed value in table \eqref{tab:constraints}.  
\\
The Hubble parameter lies within the range of $[0.913-3.04]\times10^{-10}$, whereas the temperature values at the crossing horizon fall within the range of $[3.96-1.39]\times10^{-8}$. It is evident that the condition $T > H$ is satisfied. Additionally, the evolution of both quantities is tracked in order to guarantee that the condition is satisfied in the whole inflationary epoch, and we found that $T_{ned}/H_{end} \in [85.2-12.9\times10^2]$.
\\When we calculated the number of e-foldings, we found that $N \approx 40, 39$ corresponded to values $\xi=6\times 10^5$ and $\xi=7\times 10^5$. On the other hand, the integral failed to converge to the prescribed accuracy for specific values between $\phi_\ast$ and $\phi_{end}$ for values greater than $\xi > 7\times 10^5$. 
\\The values of $Z*Q$ during the horizon-crossing era fall within $[5.68-2.70]>1$. These $Z*Q$values, however, do not provide a firm condition for a strong dissipation limit $Z*Q\gg1$. Nevertheless, as $Z*Q$ is a dynamic quantity, we observe an increase in dissipation during the inflationary period, and the values of $Z*Q$ at the end of inflation fall within $[59.1-89.9]$, excluding the first two points of the $\xi$ range in which  $Z*Q<1$. 
\\To ensure accuracy, we recalculate the observable quantities using the detailed equation in Appendix \ref{AppendixA}, rather than the approximate equations. In Table \eqref{Table_strong_points}, the values determined from the detailed equations are compared to those estimated from the approximate equation for specific benchmark points. detailed equations were used to generate all the curves in Figure \eqref{fig: strong_countors} as well as the intervals indicated previously.

In a similar way, we examine the effect of $\Upsilon_0$ on the model, where it is scanned in the range $[1-5]\times 10^5$, with the remaining parameters fixed as:
\begin{equation}\label{eq:upsilon_scan}
\xi=6\times 10^5,\lambda=1\times 10^{-7},V_0=1\times10^{-12}, \text{and} \phi=5.9.
\end{equation}
In Figure \eqref{fig: strong_countors}, the black curve indicates that increasing $\Upsilon_0$ values correspond to increasing $n_s$ in the range $[0.956-0.976]$, and decreasing $r$ in the range $[6.10-1.78]\times10^{-12}$.
\\ Spectrum amplitude depends directly on $\Upsilon_0$ and lies within $[3.07-10.5]\times10^{-9}$. Number of e-foldings falls within the range [21.7-44.1]. 
\\To ensure that $T>H$ is achieved, the values of $T$ and $H$ are checked, and we find that $H \in [3.0429-3.043]\times 10^{-10}$,and $T \in [4.55-3.835]\times10^{-8}$, and additionally, $T_{end}/H_{end} \in [126-79.7]$.Scanning shows a weak dependence of the Hubble parameter on the $Upsilon_0$. 
\\Finally, $Z*Q$ falls within the range of $[3.24-6.45]$, but unlike the $\xi$ scanning case, $(Z*Q)_{end}<1$. Once again, all calculations are made using detailed equations.  

By scanning the inflaton field within the range $[4-5.6]$ while maintaining fixed parameters as:
\begin{equation}\label{eq:phi_scan}
\xi=6\times10^5,\lambda=1\times10^{-7},\text{and} V_0=1\times10^{-12}
\end{equation}
the following observations are made:
\begin{itemize}
    \item The spectral index, $n_s$, experiences an increase within the range of $ns\in[0.957-0.9757]$.
    \item The tensor-to-scalar ratio, $r$, decreases within the range of $r\in[3.19-1.62]\times10^{-12}$.
    \item The amplitude of scalar perturbations, $A_s$, shows an increase within the range of $[5.88-11.5]\times10^{-9}$.
    \item The number of e-folds falls within the range of $[22.3-43.7]$.
    \item The condition $T > H$ is met, with $T/H$ ranging from $[158.1-127.6]$at the crossing horizon, while $T_{end}/H_{end}=75.5$. Since the scanning is performed in terms of $\phi$, one can expect the inflaton field to have the same value at the end of inflation, resulting in a constant value for the ratio at the end of inflation.
    \item $Z*Q\approx 7$, which is greater than 1, while $(Z*Q)_{end}=0.3<1$.
\end{itemize}
Finally, According to the energy scale $\lambda \in [3-9]\times 10^{-7}$, we found the following: 
\begin{itemize}
    \item The spectral index, $n_s$, increases within the range of $n_s \in [0.970 - 0.975]$.
    \item The tensor-to-scalar ratio, $r$, decreases within the range of $r \in [1.82 - 1.79] \times 10^{-12}$.
    \item The amplitude of perturbations, $A_s$, increases within the range of $[3.09 - 9.42] \times 10^{-9}$.
    \item The number of e-folds $N$ is only applicable for$ \lambda=(0.8,0.9)\times10^{-7}$ as, $N=(43.6, 45.6)$.
    \item The condition$ T > H$ holds true, with $T/H$ ranging from $[146.9 - 129.0]$ at the crossing horizon.
    \item $Z*Q$ is within the range of $[3.07 - 6.2]$. Meanwhile, $(ZQ)_{end}$ corresponds to $(122.7, 109.0)$ for $\lambda=(0.8,0.9)*10^{-7}$ respectively.
\end{itemize}
\begin{table}[]
\renewcommand{\arraystretch}{1.5}
\begin{tabular}{|l|l|l|l|l|}
\hline
             & P1      & P2      & P3      & P4      \\ \hline
$n_s$         & $0.972$ &    $0.975564$     &   $0.957443$      &     $0.966464$
    \\ \hline
$n_s(Strong)$ & $0.9683 $ &     $0.9721$    &    $0.95046$     &     $0.948071$    \\ \hline
$r$            &    $2.266\times10^{-12} $    &    $ 3.19\times10^{-12} $    &    $2.266\times10^{-12} $         &    $6.7155\times10^{-12} $  \\ \hline
$r(Strong)  $  &  $3.926\times10^{-12} $&    $2.947\times10^{-12} $     &   $5.119\times10^{-12}$      &    $1.707\times10^{-12}$     \\ \hline
$T$            &   $3.959\times10^{-8}$     &    $3.835\times10^{-8}$     &   $4.811\times10^{-8}$      &  $1.385 \times10^{-8}$      \\ \hline
$H $           &     $3.043\times10^{-10}$   &    $3.043\times10^{-10}$     &   $3.0429\times10^{-10}$       &  $ 9.12871\times10^{-11}$      \\ \hline
$As$           &    $8.282\times10^{-9}$      &    $10.52\times10^{-9}$     &   $5.87821*\times10^{-9}$      &  $ 2.51461 \times10^{-9}$     \\ \hline
$N $           &       $39.68$  &    $44.127$     &   $22.25$      &   NA      \\ \hline
$Z*Q$          &     $ 5.68$ &     $6.453$    &   $ 7.02$     &     $2.70366$
    \\ \hline
$T\_end$       &     $2.593 \times10^{-8}$   &   $2.426\times 10^{-8}$      &     $2.296\times 10^{-8}$    &     $1.17754\times 10^{-7}$     \\ \hline
$H\_end$       &      $3.0429\times10^{-10}$   &   $3.043\times10^{-10}$      &   $3.0429\times10^{-10}$      &   $9.12872\times10^{-11}$      \\ \hline
\end{tabular}
\caption{Benchmark points in the case of strong dissipation limit for the case of dissipation factor $\Upsilon=\Upsilon_0 T^3$. The parameters for the first point can be found in equation \eqref{Eq:xi_scane} with a value of $\xi=6\times 10^5$. For the second point, the dissipation factor is set to $\Upsilon_0 = 5\times10^6$, while the remaining parameters can be obtained from equation \eqref{eq:upsilon_scan}. The third point's parameters are given by equation \eqref{eq:phi_scan} with $\phi = 4.0$. Lastly, the fourth point's parameters can be found in equation \eqref{Eq:xi_scane} with $\xi = 20.0\times10^5$.}
\label{Table_strong_points}
\end{table}
\subsubsection{Weak Dissipation Limit:}
\begin{figure*}[t!]
  \begin{subfigure}{\linewidth}
  \includegraphics[scale=0.5]{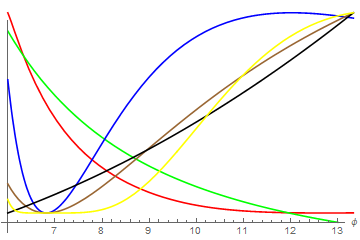}{(a)}\hfill
  \includegraphics[width=.45\linewidth]{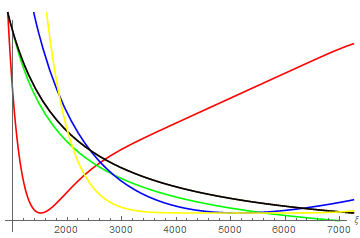}{(b)}\hfill
  \end{subfigure}\par\medskip
  \begin{subfigure}{\linewidth}
  \includegraphics[width=.45\linewidth]{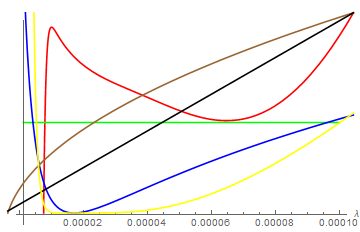}{(c)}\hfill
  \includegraphics[width=.45\linewidth]{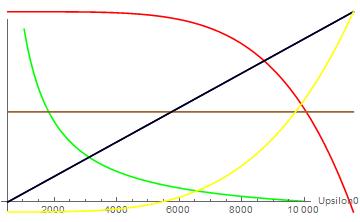}{(d)}\hfill
  \caption{}
  \end{subfigure}\par\medskip
  \begin{subfigure}{\linewidth}
  \includegraphics[width=.45\linewidth]{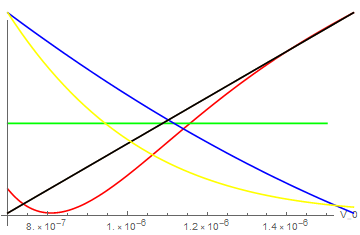}{(e)}\hfill
  \includegraphics[width=.25\linewidth]{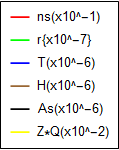}{(f)}\hfill
  \caption{}
  \end{subfigure}
  \caption{The impact of parameters $\xi$, $\lambda$, $V_0$, $\phi$, and $\Upsilon_0$ on the observables $ns$, $r$, and $As$, as well as the quantities $T$, $H$, and $Z.Q$.}\label{fig:weak_parameters_effect}
\end{figure*}

\begin{figure}
    \centering
    \includegraphics[width=1.0\linewidth]{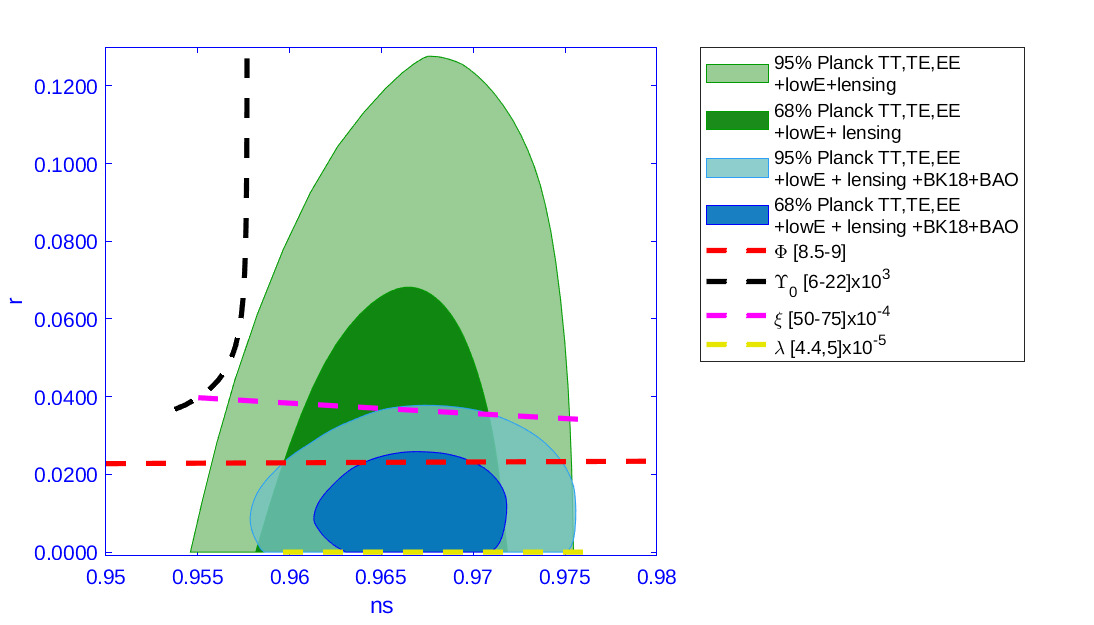}
    \caption{A plot of $n_s$ and $r$ under the weak dissipation limit. the parameter space for each represented case can be found in table \eqref{tab:weak_case_table}.}
    \label{fig:weak_countors}
\end{figure}

A similar analysis is conducted when the weak dissipation limit is considered. Analyzing this case was more challenging than analyzing the strong dissipation case. We examine the impact of individual parameters on the overall range of observations and constraints in Figure \eqref{fig:weak_parameters_effect} to gain a better understanding and identify intersections with observational results.
\\The weak limit presents particular difficulties since most coincidences with observed results are obtained with parameters that cannot lead to an end-of-inflation scenario. 
\\Even when the end of inflation is achieved, the number of e-foldings tends to be small. For example, when considering parameters as: $\xi=1.9\times 10^3$, $\Upsilon_0=5.0\times10^3$, $\lambda=5.0\times10^{-5}$, , $V_0=1.0\times10^{-6}$, and $\phi=10.0$, the resulting number of e-foldings $N$ is $20.73$. This corresponds to the observables $n_s=0.9596$, $r=1.16\times10^{-6}$, and $A_s=8.05\times10^{-7}$.
\\Upon scanning, it is observed that $T=2.20\times10^{-6}$ and $H=2.15\times10^{-6}$ are in the same order. The evolution of these dynamic quantities, however, reveals that $T<H$ at the end of inflation. When $\lambda$ is adjusted to $4.4\times10^{-5}$ while keeping other parameters fixed, the number of e-foldings increases to $N=25$. However, the challenges associated with the other constraints remain the same.
\\In Table \eqref{tab:weak_case_table}, we present the ranges of the observables $n_s$, $r$, and $A_s$ along with the important quantities $T$,$H$, and $Z.Q$ in this limit of dissipation. In conjunction with the table, we show the observables $n_s$ and $r$ compared with the experimental data of Planck 2018 and BK18 in figure \eqref{fig:weak_countors}.
\begin{table}[]
\renewcommand{\arraystretch}{1.5}
\begin{tabular}{|l|l|l|l|}
\hline
      & $\Upsilon_0=4.0*10^4$,$\lambda=23.0*10^{-14}$ & $\xi=50*10^{-4}$,$\lambda=25.0*10^{-14}$         & $\Upsilon_0=20.0*10^3$,$\lambda=25.0*10^{-14}$ \\
      & $V_0=9.0*10^{-10}$,$\xi=37*10^{-4}$           & $V_0=8.0*10^{-10}$,$\phi=10.0$                   & $V_0=8.0*10^{-10}$,$\phi=10.0$                 \\
      & $\phi=[8.5-9]$                                & $\Upsilon_0=[6-22]\times 10^{3}$                 & $\xi=[50-75]\times 10^{-4}$                    \\ \hline
$n_s$ & $[0.9809-0.9476]$                             & $[0.958-0.953]$                                  & $[0.9550-0.9764]$                              \\ \hline
$r$   & $[0.0234-0.0227]$                             & $[0.127-0.0361]$                                 & $[0.0397-0.0340]$                              \\ \hline
$T$   & $[1.29-1.52]\times 10^{-4}$                   & $[2.87-1.05]\times 10^{-5}$                      & $[9.55-6.70]\times10^{-5}$                     \\ \hline
$H$   & $[2.09-2.18]\times 10^{-5}$                   & $2.21 \times 10^{-5}$                            & $[2.21-1.89]\times 10^{-5}$                    \\ \hline
$Z.Q$ & $[1.73-2.79]\times10^{-3}$                           & $[3.19*10^{-6}-5.76*10^{-4}]$ & $[3.94-1.85]\times10^{-4}$                     \\ \hline
$A_s$ & $[3.773-4.240]\times 10^{-9}$                 & $[7.79-2.74]\times 10^{-9}$                      & $[2.49-2.14]\times 10^{-9}$                    \\ \hline
\end{tabular}
\caption{the parameter spaces' range within the weak dissipation limit, showcasing the observables $n_s$, $r$, and $A_s$, as well as key quantities $T$, $H$, and $Z.Q$.}
\label{tab:weak_case_table}
\end{table}
\section{Summery and Conclusion:}\label{sect:conclusion}
In this paper, we present a general framework for studying the warm inflationary scenario using the formalism of affine gravity. An analysis of the non-minimal coupling of the inflaton field to gravity is presented using the term $\xi\Re\phi^2$. We formulate equations of motion assuming continuous dissipation of the inflaton field into radiation through the term $\Upsilon u^\nu \nabla_\nu \phi \nabla_\mu \phi$. 
\\The warm scenario in affine gravity is considered using the slow roll approximation. We derive observable formulas for the spectral index, scalar-to-tensor ratio, and scalar amplitude under both strong and weak dissipation limits. The study provides general treatment to study different single-field potentials inspired by various elementary particle theories.
\\The following constraints are discussed:  
\begin{itemize}
    \item Temperature constraints: In contrast to cold inflation, warm inflation would require temperatures that surpass the Hubble parameter during the inflationary period. In light of the dynamic nature of temperature and Hubble parameter, it's not sufficient to achieve the constraints within the crossing horizon, but it's important to follow up throughout the inflationary period. As we studied the weak dissipation limit with quartic potential, we found that the temperature increased slower in the inflationary epoch than the Hubble parameter, so even though T was greater than H at crossing the horizon, it failed later.
    \item Strong/Weak limits constraints: The dynamics of $ZQ$ are important throughout the inflationary period as well.
    \item Number of e-foldings: In order to solve the flatness problem, N should satisfy the lower bound at least.
\end{itemize}
As well as observing the important observables of $n_s$, $r$, and $A_s$, all of the above constraints are discussed and considered extensively throughout the study of the quartic potential. 

As a result of analyzing the quartic potential, significant agreement is found with both Planck's and $BK18$ observations. 
\\Due to the modest value of the dissipation constraint within the strong dissipation limit where, $Z.Q \sim 10^1$, all results have been re-examined using detailed equations that are applicable for studying the model in general, regardless of the specific limit.
\\With a weak dissipation limit, the model achieved a coincidence with observations, but not all constraints. 
\appendix
\section{General Equations of observables:}\label{AppendixA}
 This appendix summarizes the detailed equation for $n_s$, $r$, and $A_s$.
 \\The spectral index $n_s$ shown as: 
\begin{equation}
\begin{split}
n_s-1=\frac{1}{(1+Z.Q)}\Bigg\{-\frac{\omega}{1+\omega}{\left(\frac{1}{4} (\beta+2 \eta+2\sigma-7 \varepsilon)+\frac{((3+4 \pi ) Z.Q+6) (\beta+\sigma-\varepsilon)}{(1+Z.Q) (3+4 \pi  Z.Q)}\right)}
\\ 
+2 \left(\frac{\beta+\sigma-\varepsilon}{(1+Z.Q)}-\beta+\eta-\sigma\right)-4 \varepsilon\Bigg\}
\end{split}
\end{equation}
where:
\begin{equation}
\omega=\frac{T \left(2 \sqrt{3} \pi  Q Z\right)}{H \sqrt{4 \pi  Q Z+3}}
\end{equation}
Tensor-to-scalar ratio is given by:
\begin{equation}
r=\frac{16 \varepsilon}{(2 \nu+\omega+1) (1+Z.Q)^2}
\end{equation}
where:
\begin{equation}
\nu=\frac{1}{\exp \left(\frac{H}{T}\right)-1}
\end{equation}
Scalar perturbations spectrum: 
\begin{equation}
A_s=\frac{H^4 (2 \nu+\omega+1)}{4 \pi ^2 \dot{\chi}^2}
\end{equation}



\begin{thebibliography}{99}
\bibitem{Weyl:1919fi}
H.~Weyl,
Annalen Phys. \textbf{59}, 101-133 (1919)
doi:10.1002/andp.19193641002
\bibitem{Ferraris:1982}
Ferraris, M., Francaviglia, M. and Reina, C. Variational formulation of general relativity from 1915 to 1925 “Palatini's method” discovered by Einstein in 1925. Gen Relat Gravit 14, 243–254 (1982). https://doi.org/10.1007/BF00756060
\bibitem{Palatini:1919}
Palatini, A. Deduzione invariantiva delle equazioni gravitazionali dal principio di Hamilton. Rend. Circ. Matem. Palermo 43, 203–212 (1919). https://doi.org/10.1007/BF03014670

\bibitem{Wald:1984rg}
R.~M.~Wald,
Chicago Univ. Pr., 1984,
doi:10.7208/chicago/9780226870373.001.0001
\bibitem{Ferraris:1992dx}
M.~Ferraris, M.~Francaviglia and I.~Volovich,
Class. Quant. Grav. \textbf{11}, 1505-1517 (1994)
doi:10.1088/0264-9381/11/6/015
[arXiv:gr-qc/9303007 [gr-qc]].
\bibitem{Magnano:1993bd}
G.~Magnano and L.~M.~Sokolowski,
Phys. Rev. D \textbf{50}, 5039-5059 (1994)
doi:10.1103/PhysRevD.50.5039
[arXiv:gr-qc/9312008 [gr-qc]].

\bibitem{Azri:2015diq}
H.~Azri,
Annalen Phys. \textbf{528} (2016), 404-411
doi:10.1002/andp.201500270
[arXiv:1511.06600 [gr-qc]].
\bibitem{Azri:2015vta}
H.~Azri,
Class. Quant. Grav. \textbf{32} (2015) no.6, 065009
doi:10.1088/0264-9381/32/6/065009
[arXiv:1501.06177 [gr-qc]].
\bibitem{Azri:2018qcc}
H.~Azri,
Class. Quant. Grav. \textbf{36} (2019) no.16, 165006
doi:10.1088/1361-6382/ab2e1d
[arXiv:1808.09348 [gr-qc]].



\bibitem{Azri:2018xap}
H.~Azr\i{},
[arXiv:1805.03936 [gr-qc]].
\bibitem{Azri:2020bzl}
H.~Azri, A.~Jueid, C.~Karahan and S.~Nasri,
Phys. Rev. D \textbf{102} (2020) no.8, 084036
doi:10.1103/PhysRevD.102.084036
[arXiv:2007.09681 [hep-ph]].
\bibitem{Azri:2020agc}
H.~Azri and S.~Nasri,
Phys. Rev. D \textbf{103} (2021) no.2, 024035
doi:10.1103/PhysRevD.103.024035
[arXiv:2012.04694 [gr-qc]].
\bibitem{Azri:2021dbr}
H.~Azri and S.~Nasri,
Phys. Rev. D \textbf{104} (2021) no.6, 064028
doi:10.1103/PhysRevD.104.064028
[arXiv:2107.12953 [gr-qc]].


\bibitem{Azri:2017uor}
H.~Azri and D.~Demir,
Phys. Rev. D \textbf{95} (2017) no.12, 124007
doi:10.1103/PhysRevD.95.124007
[arXiv:1705.05822 [gr-qc]].
\bibitem{Azri:2018qux}
H.~Azri and D.~Demir,
Phys. Rev. D \textbf{97} (2018) no.4, 044025
doi:10.1103/PhysRevD.97.044025
[arXiv:1802.00590 [gr-qc]].
\bibitem{Azri:2018gsz}
H.~Azri,
Int. J. Mod. Phys. D \textbf{27} (2018) no.09, 1830006
doi:10.1142/S0218271818300069
[arXiv:1802.01247 [gr-qc]].
\bibitem{Azri:2019ffj}
H.~Azri and S.~Nasri,
Phys. Rev. D \textbf{101} (2020) no.6, 064073
doi:10.1103/PhysRevD.101.064073
[arXiv:1911.11495 [gr-qc]].
\bibitem{Azri:2021uat}
H.~Azri, I.~Bamwidhi and S.~Nasri,
Phys. Rev. D \textbf{104} (2021) no.10, 104064
doi:10.1103/PhysRevD.104.104064
[arXiv:2111.03828 [gr-qc]].

\bibitem{Berera:1995ie}
A.~Berera,
Phys. Rev. Lett. \textbf{75}, 3218-3221 (1995)
doi:10.1103/PhysRevLett.75.3218
[arXiv:astro-ph/9509049 [astro-ph]].
\bibitem{Berera:1995wh}
A.~Berera and L.~Z.~Fang,
Phys. Rev. Lett. \textbf{74}, 1912-1915 (1995)
doi:10.1103/PhysRevLett.74.1912
[arXiv:astro-ph/9501024 [astro-ph]].


\bibitem{AlHallak:2021hwb}
M.~AlHallak, A.~AlRakik, N.~Chamoun and M.~S.~El-Daher,
Universe \textbf{8}, no.2, 126 (2022)
doi:10.3390/universe8020126
[arXiv:2111.05075 [astro-ph.CO]].
\bibitem{AlHallak:2022haa}
M.~AlHallak, K.~K.~A.~Said, N.~Chamoun and M.~S.~El-Daher,
Universe \textbf{9}, no.2, 80 (2023)
doi:10.3390/universe9020080
[arXiv:2211.07775 [gr-qc]].
\bibitem{Alhallak:2022szt}
M.~Alhallak, N.~Chamoun and M.~S.~Eldaher,
Eur. Phys. J. C \textbf{83}, no.6, 533 (2023)
doi:10.1140/epjc/s10052-023-11667-9
[arXiv:2212.04935 [astro-ph.CO]].
\bibitem{Samart:2021hgt}
D.~Samart, P.~Ma-adlerd, P.~Koad and P.~Channuie,
Eur. Phys. J. C \textbf{82}, no.6, 504 (2022)
doi:10.1140/epjc/s10052-022-10456-0
[arXiv:2109.09153 [astro-ph.CO]].
\bibitem{Kamali:2018ylz}
V.~Kamali,
Eur. Phys. J. C \textbf{78}, no.11, 975 (2018)
doi:10.1140/epjc/s10052-018-6449-x
[arXiv:1811.10905 [gr-qc]].
\bibitem{Bezrukov:2007ep}
F.~L.~Bezrukov and M.~Shaposhnikov,
Phys. Lett. B \textbf{659}, 703-706 (2008)
doi:10.1016/j.physletb.2007.11.072
[arXiv:0710.3755 [hep-th]].
\bibitem{AlHallak:2016wsa}
M.~AlHallak and N.~Chamoun,
JCAP \textbf{09}, 006 (2016)
doi:10.1088/1475-7516/2016/09/006
[arXiv:1604.03966 [hep-ph]].
\bibitem{Ferreira:2018nav}
R.~Z.~Ferreira, A.~Notari and G.~Simeon,
JCAP \textbf{11}, 021 (2018)
doi:10.1088/1475-7516/2018/11/021
[arXiv:1806.05511 [astro-ph.CO]].
\bibitem{AlHallak:2022gbv}
M.~AlHallak, N.~Chamoun and M.~S.~Eldaher,
JCAP \textbf{10}, 001 (2022)
doi:10.1088/1475-7516/2022/10/001
[arXiv:2202.01002 [astro-ph.CO]].
\bibitem{AlHallak:2023wsx}
M.~AlHallak,
Phys. Sci. Forum \textbf{7}, no.1, 35 (2023)
doi:10.3390/ECU2023-14048
[arXiv:2302.06524 [astro-ph.CO]].

\bibitem{AlHallak:2021sic}
M.~AlHallak, A.~Al Rakik, S.~Bitar, N.~Chamoun and M.~S.~Eldaher,
Int. J. Mod. Phys. A \textbf{36}, no.30, 2150226 (2021)
doi:10.1142/S0217751X21502262
[arXiv:2105.00848 [astro-ph.CO]].
\bibitem{Clarkson:2003af}
C.~A.~Clarkson,
Phys. Rev. D \textbf{70}, 103524 (2004)
[erratum: Phys. Rev. D \textbf{70}, 129902 (2004)]
doi:10.1103/PhysRevD.70.129902
[arXiv:astro-ph/0311505 [astro-ph]].
\bibitem{Liddle:2000cg}
 A. R. Liddle and D. H. Lyth, Cosmological Inflation and LargeScale Structure (Cambridge University Press, Cambridge, England, 2000).
\bibitem{Weinberg:2008zzc}
S. Weinberg, Cosmology (Oxford University Press, Oxford,
England, 2008).
\bibitem{Berera:1999ws}
A.~Berera,
Nucl. Phys. B \textbf{585}, 666-714 (2000)
doi:10.1016/S0550-3213(00)00411-9
[arXiv:hep-ph/9904409 [hep-ph]].




\bibitem{Albrecht:1982mp}
A.~Albrecht, P.~J.~Steinhardt, M.~S.~Turner and F.~Wilczek,
``Reheating an Inflationary Universe,''
Phys. Rev. Lett. \textbf{48}, 1437 (1982)
doi:10.1103/PhysRevLett.48.1437
\bibitem{Hall:2004zr}
L.~M.~H.~Hall and I.~G.~Moss,
``Thermal effects on pure and hybrid inflation,''
Phys. Rev. D \textbf{71}, 023514 (2005)
doi:10.1103/PhysRevD.71.023514
[arXiv:hep-ph/0408323 [hep-ph]].
\bibitem{Hall:2003zp}
L.~M.~H.~Hall, I.~G.~Moss and A.~Berera,
``Scalar perturbation spectra from warm inflation,''
Phys. Rev. D \textbf{69}, 083525 (2004)
doi:10.1103/PhysRevD.69.083525
[arXiv:astro-ph/0305015 [astro-ph]].
\bibitem{Abbott:1982hn}
L.~F.~Abbott, E.~Farhi and M.~B.~Wise,
``Particle Production in the New Inflationary Cosmology,''
Phys. Lett. B \textbf{117}, 29 (1982)
doi:10.1016/0370-2693(82)90867-X


\bibitem{Manton:1983nd}
N.~S.~Manton,
Phys. Rev. D \textbf{28}, 2019 (1983)
doi:10.1103/PhysRevD.28.2019
\bibitem{Arnold:1987mh}
P.~B.~Arnold and L.~D.~McLerran,
Phys. Rev. D \textbf{36}, 581 (1987)
doi:10.1103/PhysRevD.36.581
\bibitem{Arnold:1987zg}
P.~B.~Arnold and L.~D.~McLerran,
Phys. Rev. D \textbf{37}, 1020 (1988)
doi:10.1103/PhysRevD.37.1020



\bibitem{Mirbabayi:2022cbt}
M.~Mirbabayi and A.~Gruzinov,
JCAP \textbf{02} (2023), 012
doi:10.1088/1475-7516/2023/02/012
[arXiv:2205.13227 [astro-ph.CO]].
\bibitem{Arnold:1996dy}
P.~B.~Arnold, D.~Son and L.~G.~Yaffe,
Phys. Rev. D \textbf{55}, 6264-6273 (1997)
doi:10.1103/PhysRevD.55.6264
[arXiv:hep-ph/9609481 [hep-ph]].
\bibitem{Bastero-Gil:2016qru}
M.~Bastero-Gil, A.~Berera, R.~O.~Ramos and J.~G.~Rosa,
Phys. Rev. Lett. \textbf{117} (2016) no.15, 151301
doi:10.1103/PhysRevLett.117.151301
[arXiv:1604.08838 [hep-ph]].


\bibitem{Planck:2018jri}
Y.~Akrami \textit{et al.} [Planck],
Astron. Astrophys. \textbf{641}, A10 (2020)
doi:10.1051/0004-6361/201833887
[arXiv:1807.06211 [astro-ph.CO]].
\bibitem{Planck:2013pxb}
P.~A.~R.~Ade \textit{et al.} [Planck],
Astron. Astrophys. \textbf{571}, A16 (2014)
doi:10.1051/0004-6361/201321591
[arXiv:1303.5076 [astro-ph.CO]].
\end{thebibliography}
\end{document}